\documentclass[12pt]{article}
\usepackage{epsfig,amsbsy}
\newtheorem{theorem}{Theorem}[section]

\newtheorem{definition}[theorem]{Definition}

\makeatletter
\@addtoreset{equation}{section}
\makeatother

\newcommand{\ret}{\nonumber \\}

\newcommand{\abs}[1]{\left|#1\right|}

\newcommand{\rbk}[1]{\left(#1\right)}
\newcommand{\sqbk}[1]{\left[#1\right]}
\newcommand{\cbk}[1]{\left\{#1\right\}}
\newcommand{\bkt}[1]{\left\langle#1\right\rangle}

\newcommand{\sumtwo}[2]%
{\mathop{\sum_{#1}}_{#2}}
\newcommand{\sumthree}[3]%
{\mathop{\mathop{\sum_{#1}}_{#2}}_{#3}}
\newcommand{\sumfour}[4]%
{\mathop{\mathop{\mathop{\sum_{#1}}_{#2}}_{#3}}_{#4}} 
\newcommand{\suptwo}[2]%
{\mathop{\sup_{#1}}_{#2}}
\newcommand{\supthree}[3]%
{\mathop{\mathop{\sup_{#1}}_{#2}}_{#3}}
\newcommand{\supfour}[4]%
{\mathop{\mathop{\mathop{\sup_{#1}}_{#2}}_{#3}}_{#4}} 
\newcommand{\inftwo}[2]%
{\mathop{\inf_{#1}}_{#2}}
\newcommand{\infthree}[3]%
{\mathop{\mathop{\inf_{#1}}_{#2}}_{#3}}
\newcommand{\inffour}[4]%
{\mathop{\mathop{\mathop{\inf_{#1}}_{#2}}_{#3}}_{#4}} 
\newcommand{\La}{\Lambda}
\newcommand{\up}{\uparrow}
\newcommand{\dn}{\downarrow}

\newcommand{\cxu}{c^\dagger_{x,\up}}
\newcommand{\cxd}{c^\dagger_{x,\dn}}
\newcommand{\axu}{c_{x,\up}}
\newcommand{\axd}{c_{x,\dn}}
\newcommand{\cxs}{c^\dagger_{x,\sigma}}
\newcommand{\axs}{c_{x,\sigma}}

\newcommand{\ays}{c_{y,\sigma}}
\newcommand{\cyt}{c^\dagger_{y,\tau}}
\newcommand{\ayt}{c_{y,\tau}}
\newcommand{\cd}{c^\dagger}
\newcommand{\nxs}{n_{x,\sigma}}
\newcommand{\nxu}{n_{x,\up}}
\newcommand{\nxd}{n_{x,\dn}}
\newcommand{\Emin}{E_{\rm min}}

\newcommand{\Hhop}{H_{\rm hop}}
\newcommand{\Hint}{H_{\rm int}}
\newcommand{\Stot}{S_{\rm tot}}
\newcommand{\Sztot}{S_{\rm tot}^{(3)}}
\newcommand{\Smax}{S_{\rm max}}
\newcommand{\vac}{\Phi_{\rm vac}}
\newcommand{\GS}{\Phi_{\rm GS}}

\newcommand{\Neop}{\hat{N}_{\rm e}}
\newcommand{\Ne}{N_{\rm e}}
\newcommand{\Sop}{\hat{\bf S}}
\newcommand{\Sopc}[2]{\hat{S}^{(#1)}_{#2}}
\newcommand{\Hub}{Hubbard model}
\newcommand{\Sch}{Schr\"{o}dinger}
\renewcommand{\phi}{\varphi}
\newcommand{\ep}{\varepsilon}

\newcommand{\tsigma}{\boldsymbol{\sigma}}
\newcommand{\ket}[1]{{\left|#1\right\rangle}}
\newcommand{\tHhop}{\widetilde{H}_{\rm hop}}
\newcommand{\Ns}{|\La|}
\newcommand{\vphi}{\boldsymbol{\phi}}
\begin{document}
%%%%%%%%%%%%%%%%%%%%%%%%%%
%%%%%%%%%%%%%%%%%%%%%%%%%%
%%%%%%%%%%%%%%%%%%%%%%%%%%
%%%%%%%%%%%%%%%%%%%%%%%%%%
\begin{center}
{\bf
The Hubbard Model
---Introduction and Selected Rigorous Results}
\par\bigskip
Hal Tasaki\footnote{
hal.tasaki@gakushuin.ac.jp,
http://www.gakushuin.ac.jp/\( \tilde{\ } \)881791/
%% it's  http://www.gakushuin.ac.jp/~881791/
}
\par\bigskip
{\footnotesize\sl Department of Physics, Gakushuin University,
Mejiro, Toshima-ku, Tokyo 171, JAPAN}
\end{center}
%%%%%%%%%%%%%%%%%%%%%%%%%%%%%%%%%%%%%%
%\vfil
%%%%%%%%%%%%%%%%%%%%%%%%%%%%%%%%%%%%%%
\begin{abstract}
The Hubbard model is a ``highly oversimplified model'' for electrons 
in a solid which interact with each other through extremely short 
ranged repulsive (Coulomb) interaction.
The Hamiltonian of the Hubbard model consists of two pieces; 
$\Hhop$ which describes quantum mechanical hopping of electrons, 
and $\Hint$ which describes nonlinear repulsive interaction.
Either $\Hhop$ or $\Hint$ alone is easy to analyze, and 
does not favor any specific order.
But their sum $H=\Hhop+\Hint$ is believed to exhibit various nontrivial 
phenomena including metal-insulator transition,
antiferromagnetism, ferrimagnetism, ferromagnetism, 
Tomonaga-Luttinger liquid, and superconductivity.
It is believed that we can find various interesting ``universality 
classes'' of strongly interacting electron systems by studying the 
idealized Hubbard model.

In the present article we review some mathematically rigorous results on 
the Hubbard model which shed light on ``physics'' of this fascinating 
model. 
We mainly 
concentrate on magnetic properties of the model at its ground states.
We discuss Lieb-Mattis theorem on the absence of ferromagnetism in one 
dimension, 
Koma-Tasaki bounds on decay of 
correlations at finite temperatures in two-dimensions,
Yamanaka-Oshikawa-Affleck theorem on low-lying 
excitations in one-dimension, 
Lieb's important theorem for half-filled model on a bipartite 
lattice, 
Kubo-Kishi bounds on the charge and superconducting susceptibilities of 
half-filled models at finite temperatures,
and three rigorous examples of saturated ferromagnetism due to 
Nagaoka, Mielke, and Tasaki.
We have tried to make the article accessible to nonexperts by  
describing basic definitions and elementary materials in detail.
\end{abstract}
%%%%%%%%%%%%%%%%%%%%%%%%%%%%%%%%%%%%%%
%\vfil
%%%%%%%%%%%%%%%%%%%%%%%%%%%%%%%%%%%%%%
%\newpage
\tableofcontents
%%%%%%%%%%%%%%%%%%%%%%%%%%%%%%%%%%%%%%
%%%%%%%%%%%%%%%%%%%%%%%%%%%%%%%%%%%%%%
%\newpage
\section{Introduction}
%%%%%%%%%%%%%%%%%%%%%%%%%%%%%%%%%%%%%%
\label{secAbout}
According to the textbook of Ashcroft and Mermin, the \Hub\ is 
``a highly 
over\-simplified model'' for
strongly interacting electrons in a solid.
The \Hub\ is a kind of minimum model which takes into account quantum 
mechanical motion of electrons in a solid, and nonlinear repulsive 
interaction between electrons.
There is little doubt that the model is too simple to describe actual solids 
faithfully.

Nevertheless, the \Hub\ is one of the most important models in theoretical 
physics.
In spite of its simple definition, the \Hub\ is believed to exhibit 
various interesting phenomena including metal-insulator transition,
antiferromagnetism, ferrimagnetism, ferromagnetism, 
Tomonaga-Luttinger liquid, and superconductivity.
Serious theoretical studies have also revealed that to understand 
various properties of the \Hub\ is a very difficult problem.
We believe that in course of getting deeper understanding of the \Hub, 
we will learn 
many new physical and mathematical techniques, concepts, 
and ways of thinking.
Perhaps a more important point comes from the idea of 
``universality.''
We believe that nontrivial phenomena and mechanisms found 
in the idealized Hubbard model can also be found
in other systems in the same 
``universality class'' as the idealized model.
The universality class is expected to be large and rich enough so that 
it contains various realistic strongly interacting electron
systems with complicated 
details which are ignored in the idealized model.

The situation is very similar to that of the Ising model
for classical spin systems.
The Ising model is 
too simple to be a realistic model of magnetic materials, but has 
turned out to be extremely important and useful in developing 
various notions 
and techniques in statistical physics of many degrees of freedom.
Many important universality classes (of spin systems and 
field theories) were discovered by studying the Ising model.

In the present article, we review some 
mathematically rigorous results\footnote{
In order to reduce the number of references, 
we have decided not to include many important references on the 
related topics which do not provide rigorous results.
} known for 
the \Hub.
We shall concentrate ourselves 
mainly
on magnetic properties of the model at its ground state, i.e., at the zero 
temperature.
We have also decided not to cover many 
important rigorous and/or exact results 
in one-dimensional models based on the Bethe ansatz solutions.
Even with these restrictions, we do not try to cover all the existing 
rigorous results.
We recall that there is an excellent review article by Lieb \cite{Lieb95} 
which covers wider topics than we do here.
As for the more restricted topics of Nagaoka's ferromagnetism, flat-band 
ferromagnetism, and some related topics, there is a separate review 
\cite{97c} which is more detailed and elementary than the present one.

%%%%%%%%%%%%%%%%%%%%%%%%%%%%%%%%%%%%%%
%%%%%%%%%%%%%%%%%%%%%%%%%%%%%%%%%%%%%%
\section{Hubbard Model}
%%%%%%%%%%%%%%%%%%%%%%%%%%%%%%%%%%%%%%
\subsection{Definition of the Hubbard Model}
\label{s:def}
\begin{figure}
\centerline{\epsfig{file=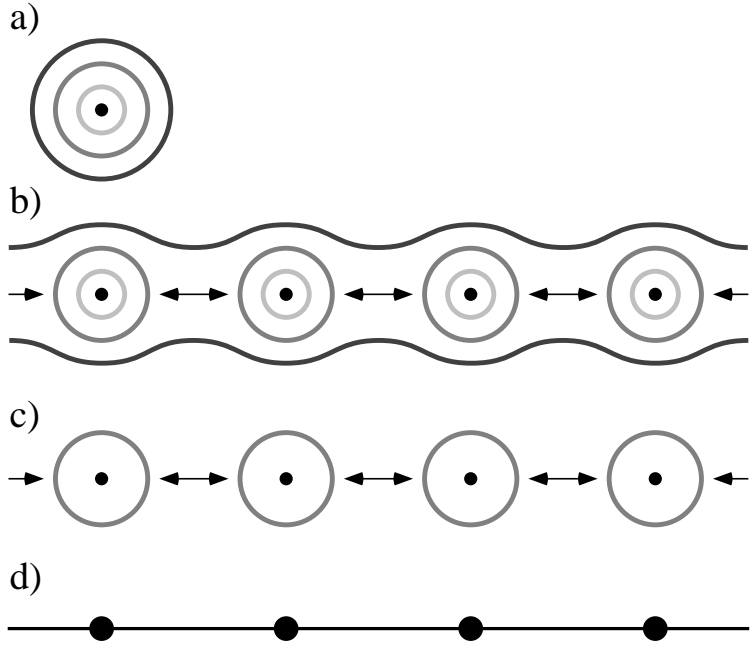,width=10cm}}
\caption[dummy]{
A highly schematic figure which explain philosophy of tight-binding 
descriptions.
a)~A single atom with multiple electrons in different orbits.
b)~When atoms get together to form a solid, electrons in the black 
orbits become itinerant, while those in the light gray orbits are 
still localized at the original atomic sites.
Electrons in the gray orbits are mostly localized around the atomic 
sites, but tunnel to nearby gray orbits with a non-negligible 
probabilities.
c)~We only consider the electrons in the gray orbits, which are 
expected to play essential roles in determining various low energy 
physics of the system.
d)~If the gray orbit is non-degenerate, we get a lattice model in 
which electrons live on lattice sites and hop from site to another.
}
\label{f:tb}
\end{figure}

We first give general definition of the \Hub\footnote{
The readers who are new to the filed are recommended to take a look 
at \cite{97c}, which contains more careful introduction to the Hubbard 
model.
}.
Let the lattice $\La$ be a collection of sites $x,y,\ldots$.
Physically speaking, each lattice site corresponds to an atomic site in a 
crystal.
In the standard Hubbard model, one simplifies the situation considerably, and 
assumes that each atom has only one electron orbit and the corresponding
orbital state is non-degenerate\footnote{
Such a model is usually referred to as a single-band Hubbard model.
This terminology is confusing since such a model can possess more than one 
(single-electron) band depending on the lattice structure.
Perhaps ``single-orbital Hubbard model'' is a better terminology.
}.
Of course actual atoms can have more then one 
orbits (or bands) and electrons in the 
corresponding states.
The philosophy behind the model building is that those electrons in other 
states do not play significant roles in low-energy physics that we are 
interested in, and can be ``forgotten'' for the moment.
See Figure~\ref{f:tb}.

By $\cxs$, we denote the operator which creates an electron with spin 
$\sigma=\up,\dn$ at site  $x\in\La$. 
The corresponding annihilation operator is  $\axs$, and $\nxs=\cxs\axs$ is 
the number operator.
These fermion operators obey the canonical anticommutation relations
\begin{equation}
\cbk{\cxs,\ayt}=\delta_{x,y}\delta_{\sigma,\tau},
\label{ac1}
\end{equation}
and
\begin{equation}
\cbk{\cxs,\cyt}=\cbk{\axs,\ayt}=0,
\label{ac2}
\end{equation}
where $\cbk{A,B}=AB+BA$.

By $\vac$ we denote the state without any electrons.
We have  $\axs\vac=0$ for any $x\in\La$ and $\sigma=\up,\dn$.
The Hilbert space of the model is generated by the states obtained by 
successively operating the creation operator  $\cxs$ with various  $x$ 
and  $\sigma$ onto the state  $\vac$.
Since the anticommutation relation (\ref{ac2}) implies $(\cxs)^2=0$, each 
lattice site can either be vacant, occupied by an $\up$ or $\dn$ electron, 
or occupied by both $\up$ and  $\dn$ electrons.
The total dimension of the Hilbert space is thus\footnote{
Throughout the present article we denote by  $|S|$ the number of elements 
in a set $S$.
} $4^{|\La|}$.

The Hamiltonian of the \Hub\ is most naturally represented as the sum of two 
terms as
\begin{equation}
H=\Hhop+\Hint.
\label{H1}
\end{equation}
The most general form of the hopping Hamiltonian  $\Hhop$ is\footnote{
The standard convention is to put a minus sign in front of the summation 
in (\ref{Hhop}), and to assume $t_{x,y}\ge0$.
However it seems that there is no simple reason that the hopping amplitude 
should have such signs.
If the system is bipartite (See Definition \ref{bipartite}), one can 
change the sign of all $t_{x,y}$ ($x\ne y$) by performing a gauge 
transformation $\cxs\to-\cxs$ for all $x\in A$.}
\begin{equation}
\Hhop = \sum_{x,y\in\La}\sum_{\sigma=\up,\dn}t_{x,y}\,\cxs\ays.
\label{Hhop}
\end{equation}
The hopping amplitude $t_{x,y}=t_{y,x}$, which is assumed to be real, 
represents the quantum mechanical
amplitude that an electron hops from site  $x$ to  $y$ (or 
from  $y$ to  $x$).
When  $x=y$, the summand in (\ref{Hhop}) becomes 
$t_{x,x}\,\cxs\axs=t_{x,x}\,\nxs$, which is nothing but a single-body 
potential.

The interaction Hamiltonian $\Hint$ is written as
\begin{equation}
\Hint=\sum_{x\in\La}U_{x}\,\nxu\nxd,
\label{Hint}
\end{equation}
where $U_{x}>0$ is a constant.
The Hamiltonian represents a nonlinear interaction which raises the energy 
by  $U_x$ when two electrons occupy a single orbital state at  $x$.
Although the original Coulomb interaction is long ranged, we have 
``oversimplified'' the situation and took into account the strongest part 
out of the interaction\footnote{
There are many important works on various extended Hubbard models in 
which one takes into account 
other short range interactions which  arise from 
the original Coulomb interaction.
See \cite{SV,BKS,Boer95,Vollhardt97} and many references therein.
}.
Another interpretation is that the Coulomb interaction is screened by the 
electrons in different orbital states which we had decided to forget.
%%%%%%%%%%%%%%%%%%%%%%%%%%%%%%%%%%%%%%
\subsection{Some Physical Quantities}
\label{s:quant}
We shall define some basic conserved quantities.
The total number operator
\begin{equation}
\Neop=\sum_{x\in\La}(\nxu+\nxd)
\label{Ne}
\end{equation} 
commutes with the Hamiltonian $H$.
Although there are some conserved quantities other than $\Neop$, one 
usually discusses stationary states or equilibrium states of the system by 
keeping the eigenvalue or the expectation value of $\Neop$ 
constant\footnote{
For example the total spin is also a conserved quantity.
But we do not fix its eigenvalue or expectation value, since the total 
spin is not definitely conserved in reality because there is an
{\em LS} coupling and the actual solids are not rotation invariant.
The situation for \( \Neop \) is essentially different since the 
charge conservation is an exact law.
See Section~2.2 of \cite{97c}.
}.
In the present article, we mostly consider\footnote{
Section~\ref{s:KK} is the only exception.
} the Hilbert space in 
which the number operator  $\Neop$ has a fixed eigenvalue  $\Ne$.
Since each lattice site can have at most two electrons, we have 
$0\le\Ne\le2|\La|$.
The total electron number  $\Ne$ is the most fundamental parameter in the 
\Hub.

The spin operator $\Sop_x=(\Sopc{1}{x},\Sopc{2}{x},\Sopc{3}{x})$ at site  
$x$ is defined as
\begin{equation}
\Sopc{\alpha}{x}=\frac{1}{2}
\sum_{\sigma,\tau=\up,\dn}
\cxs(p^{(\alpha)})_{\sigma,\tau}\,\axs,
\label{Salpha}
\end{equation}
for  $\alpha=1,2$, and $3$, where $p^{(\alpha)}$ are the Pauli matrices.
The operators for the total spin of the system are defined as
\begin{equation}
\Sopc{\alpha}{\rm tot}=\sum_{x\in\La}\Sopc{\alpha}{x},
\label{Stot}
\end{equation}
for  $\alpha=1,2$, and $3$.
The operator $\Sopc{\alpha}{\rm tot}$ commute with both the hopping 
Hamiltonian  $\Hhop$  (\ref{Hhop}) and with the interaction Hamiltonian  
$\Hint$  (\ref{Hint}).
In other words, these Hamiltonians are invariant under any global 
rotation in the spin space.

As the operators $\Sopc{\alpha}{\rm tot}$ with  $\alpha=1,2,3$ do not 
commute with each other, we follow the convention in the theory of angular 
momenta, and simultaneously diagonalize the total spin operators 
$\Sopc{3}{\rm tot}$, $(\Sop_{\rm 
tot})^2=\sum_{\alpha=1}^3(\Sopc{\alpha}{\rm tot})^2$, and the Hamiltonian 
$H$.
We denote by $\Sztot$ and $\Stot(\Stot+1)$ the eigenvalues of 
$\Sopc{3}{\rm tot}$ and $(\Sop_{\rm tot})^2$, respectively.
For a given electron number  $\Ne$, we let
\begin{equation}
\Smax=\cases{
\Ne/2& when $0\le\Ne\le|\La|$;\cr
|\La|-(\Ne/2)&when $|\La|\le\Ne\le2|\La|$.\cr
}
\label{Smax}
\end{equation}
Then the possible values of  $\Stot$ are $\Stot=0,1,\ldots,\Smax$ (or 
$\Stot=1/2,3/2,\ldots,\Smax$).

When we discuss magnetism of the system, the most important issue is to 
determine the value of  $\Stot$ in the ground state(s).
If the total spin of the ground state grows proportionally to the number 
of sites  $|\La|$ as we increase the size of  $\La$, we say that the system exhibits 
ferromagnetism in a broad sense.
This roughly means that the system behaves as a ``magnet.''
If the total spin of the ground state(s) coincides with the maximum 
possible value $\Smax$, we say that the system exhibits saturated 
ferromagnetism.

The following quantity will be useful in the later analysis.

\begin{definition}[The lowest energy for each \( \Stot \)]
Fix the electron number  $\Ne$. 
For  $S=0,1,\ldots,\Smax$ (or $S=1/2,3/2,\ldots,\Smax$),
we denote by $\Emin(S)$ the 
lowest possible energy among the states which satisfy $\hat{N}_{\rm 
e}\Phi=\Ne\Phi$ and $(\Sop_{\rm tot})^{2}\Phi=S(S+1)\Phi$ (i.e., 
\( \Stot=S \)).
\label{Emin}
\end{definition}

The appearance of saturated ferromagnetism is equivalent to have 
$\Emin(S)>\Emin(\Smax)$ for any $S$ such that $S<\Smax$.

%%%%%%%%%%%%%%%%%%%%%%%%%%%%%%%%%%%%%%
%%%%%%%%%%%%%%%%%%%%%%%%%%%%%%%%%%%%%%
\section{Basic Facts about the Model}
In order to understand the meaning of the Hamiltonian of the \Hub, we 
discuss physics we encounter in two limiting situations.
%%%%%%%%%%%%%%%%%%%%%%%%%%%%%%%%%%%%%%
\subsection{Non-Interacting System}
\label{secfree}
Let us assume that the Coulomb interaction in $\Hint$  (\ref{Hint}) 
satisfies $U_{x}=0$ for any  $x\in\La$. 
Since the remaining Hamiltonian $H=\Hhop$ (\ref{Hhop}) is a quadratic form 
in fermion operators, it can be diagonalized easily (in principle).
The single-electron \Sch\ equation corresponding to the hopping 
Hamiltonian $\Hhop$ (\ref{Hhop}) is
\begin{equation}
\sum_{y\in\La}t_{x,y}\,\phi_y=\ep\phi_x,
\label{Sch}
\end{equation}
where $\vphi=(\phi_x)_{x\in\La}$ is a single-electron wave function, and 
$\ep$ is the single-electron energy eigenvalue.
We shall denote the eigenvalues and the eigenstates of (\ref{Sch}) as 
$\ep_j$ and $\vphi^{(j)}=(\phi^{(j)}_x)_{x\in\La}$, respectively, where the 
index takes values $j=1,2,\ldots,|\La|$.
We count the energy levels with taking degeneracies into account, and 
order them as $\ep_j\le\ep_{j+1}$.

Let us discuss a simple and standard example.
Take a one dimensional lattice $\La=\cbk{1,2,\ldots,N}$, and impose a 
periodic boundary condition which identifies the site $1$ with the site 
$N+1$.
As for the hopping matrix elements, we set $t_{x,x+1}=t_{x+1,x}=-t$, 
and $t_{x,y}=0$ otherwise.
The corresponding \Sch\ equation (\ref{Sch}) can be solved easily.
By using the wave number $k=2\pi n/N$ (with 
$n=0,\pm1,\pm2,\ldots,\pm\cbk{(N/2)-1},N/2$), the eigenstates and the 
eigenvalues can be written as $N^{-1/2}\exp[ikx]$ and $\ep(k)=-2t\cos k$, 
respectively.
If one makes a suitable correspondence between  $n$ and $j=1,2,\ldots,N$, 
we get the  desired energy level $\ep_{j}$.

We return to the general setting, and define fermion operators 
corresponding to the eigenstates $\vphi^{(j)}=(\phi^{(j)}_x)_{x\in\La}$ by
\begin{equation}
a^{\dagger}_{j,\sigma}=\sum_{x\in\La}\phi^{(j)}_x 
c^{\dagger}_{x,\sigma}.
\label{a}
\end{equation}
By using the orthonormality of the set of eigenstates 
$(\vphi^{(j)})_{j=1,2,\ldots,|\La|}$ (we redefine the eigenstates if they 
do not form an orthonormal set), one finds that the inverse transformation 
of (\ref{a}) is $\axs=\sum_{j=1}^{|\La|}\phi^{(j)}_xa_{j,\sigma}$.
Substituting this into (\ref{Hhop}), and by using (\ref{Sch}), we find 
that  $\Hhop$ can be diagonalized as
\begin{equation}
\Hhop
=
\sum_{\sigma=\up,\dn}\sum_{j=1}^{|\La|}
\ep_j\,a^\dagger_{j,\sigma}a_{j,\sigma}
=
\sum_{\sigma=\up,\dn}\sum_{j=1}^{|\La|}
\ep_j\,\tilde{n}_{j,\sigma}.
\label{Hhop2}
\end{equation}
Here $\tilde{n}_{j,\sigma}=a^\dagger_{j,\sigma}a_{j,\sigma}$ can be 
interpreted as the electron number operator for the $j$-th single-electron 
eigenstate.

Let $A,B$ be two arbitrary subsets of $\cbk{1,2,\ldots,|\La|}$ which 
satisfy $|A|+|B|=\Ne$.
By using (\ref{Hhop2}), we find that the state
\begin{equation}
\Phi_{A,B}=
\rbk{\prod_{j\in A}a^\dagger_{j,\up}}
\rbk{\prod_{j\in B}a^\dagger_{j,\dn}}
\vac
\label{PhiAB}
\end{equation}
is an eigenstate of $H=\Hhop$ and its energy eigenvalue is
\begin{equation}
E_{A,B}=\sum_{j\in A}\ep_j+\sum_{j\in B}\ep_{j}.
\label{EAB}
\end{equation}
By choosing subsets $A,B$ which minimize $E_{A,B}$, we get ground state(s) 
of the non-interacting model.

\begin{figure}
\centerline{\epsfig{file=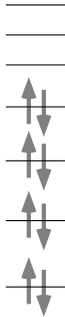,width=0.9cm}}
\caption[dummy]{
Schematic picture of the ground state of a non-interacting many 
electron system.
The lowest \( \Ne/2 \) single-electron energy levels are ``filled'' 
by both up spin and down spin electrons.
The state naturally exhibits paramagnetism known as Pauli 
paramagnetism.
}
\label{f:pauli}
\end{figure}

In particular if the corresponding single-electron energy eigenvalues are 
nondegenerate, i.e., $\ep_{j}<\ep_{j+1}$, and $\Ne$ is even, the ground 
state of $H=\Hhop$ is unique and written as
\begin{equation}
\GS=\rbk{\prod_{j=1}^{\Ne/2}a^\dagger_{j,\up}a^\dagger_{j,\dn}}\vac.
\label{GS1}
\end{equation}
This is nothing but the state obtained by ``filling up'' the low 
energy levels by up and down spin electrons, as one learns in 
elementary quantum mechanics (Figure~\ref{f:pauli}).
It is easily verified that the above state has a definite total spin 
$\Stot=0$.
The ground state (\ref{GS1}) exhibits no long range order.
A system with no magnetic ordering is usually said to exhibit 
paramagnetism\footnote{
More precisely, this is true when one talks only about magnetism carried 
by electron spins.
If one takes into account magnetism induced by orbital motion of 
electrons, the system may exhibit diamagnetism.
}.

In the simple example in one dimension, all the energy levels except 
$\ep(0)=-2t$ and $\ep(\pi)=2t$ are two-fold degenerate.
In this case the ground state of $H=\Hhop$ may not be unique for some 
values of  $\Ne$.
However the degeneracy of the ground states is at most four-fold, and the 
total spin of the ground states can take values in $\Stot=0,1/2$, or $1$.
We can conclude that the property of the ground state(s) 
is essentially the same as 
the models without degeneracy.
In general we can draw the same conclusion unless the single-electron 
spectrum has a bulk degeneracy.

In a single-electron eigenstate of the example in one dimension, the 
electron is in a plane wave state with a definite wave number  $k$. 
The fact that the Hamiltonian $H=\Hhop$ is diagonalized as (\ref{Hhop2}) 
implies that the electrons behave as ``waves'' in this non-interacting 
(Hubbard) model.
The same comment applies to any translation invariant (Hubbard) 
model with $U_{x}=0$.

%%%%%%%%%%%%%%%%%%%%%%%%%%%%%%%%%%%%%%
\subsection{Non-Hopping System}
\label{secnohop}
Let us next assume that the hopping matrix elements in $\Hhop$  
(\ref{Hhop}) satisfy $t_{x,y}=0$ for any  $x,y\in\La$. 
Then the remaining Hamiltonian $H=\Hint$ (\ref{Hint}) is already in a 
diagonal form.
A general eigenstate can be written as
\begin{equation}
\Psi_{X,Y}=
\rbk{\prod_{x\in X}c^\dagger_{x,\up}}
\rbk{\prod_{x\in Y}c^\dagger_{x,\dn}}
\vac.
\label{PsiXY}
\end{equation}
Here $X$ and $Y$ are arbitrary subsets of $\La$, and represent lattice 
sites which are occupied by up-spin electrons and down-spin electrons, 
respectively.
The total electron number in this state is $\Ne=|X|+|Y|$, and the energy 
eigenvalue is given by
\begin{equation}
E_{X,Y}=\sum_{x\in X\cap Y}U_x.
\label{EXY}
\end{equation}
The ground state for a given electron number $\Ne$ can be constructed by 
choosing subsets $X,Y$ that minimize the energy $E_{X,Y}$.
When one has $\Ne\le|\La|$, one can always choose  $X$ and $Y$ 
such that $X\cap Y=\emptyset$.
Thus the ground state has energy equal to  $0$.

The ground states of the non-hopping Hubbard model has no magnetic long 
range order.
Again the system is paramagnetic.
It is also clear (from the beginning) that the electrons behave as 
``particles'' in non-hopping models.

%%%%%%%%%%%%%%%%%%%%%%%%%%%%%%%%%%%%%%
\subsection{Hubbard Model is Difficult, but it is Interesting}
\label{secexciting}
We have investigated the properties of the two pieces  $\Hhop$ and $\Hint$ 
in the Hubbard Hamiltonian.
It turned out that both $\Hhop$ and  $\Hint$ are easy to analyze.
We also found that neither of them favor any magnetic ordering.

We also observed, however, that electrons behave as ``waves'' in $\Hhop$, 
while they behave as ``particles'' in $\Hint$.
How do they behave in a system with the Hamiltonian which is a sum of 
these totally different Hamiltonians?
This is indeed a fascinating problem which is deeply related to the 
wave-particle dualism in quantum physics.
We might say that many of the important models in many-body problems, 
including the $\phi^4$ quantum field theory and the Kondo problem, 
are minimum models which take 
into account  at the same time the wave-like nature and the particle-like 
nature (through point-like nonlinear interaction) of matter.

 From a technical point of view, the wave-particle dualism implies that 
the Hamiltonians $\Hint$ are $\Hhop$ do not commute with each other.
Even when each Hamiltonian is diagonalized, it is still highly nontrivial 
(or impossible) to find the properties of their sum.
Of course mathematical difficulty does not automatically guarantee that the 
model is worth studying.
A really exciting thing about the Hubbard model is that, though the 
Hamiltonians $\Hhop$ and $\Hint$ do not favor any nontrivial order, their 
sum  $H=\Hhop+\Hint$ is believed to generate various nontrivial order 
including antiferromagnetism, ferromagnetism, and superconductivity.
When we sum up the two innocent Hamiltonians $\Hhop$ and $\Hint$, 
competition between wave-like character and particle-like character (or 
between linearity and nonlinearity) takes place, and one gets various 
interesting ``physics.''
To confirm this fascinating scenario is a very challenging problem to 
theoretical and mathematical physicists.
%%%%%%%%%%%%%%%%%%%%%%%%%%%%%%%%%%%%%%
%%%%%%%%%%%%%%%%%%%%%%%%%%%%%%%%%%%%%%
\section{Results for Low Dimensional Models}
\label{s:lowD}
We discuss some theorems which are proved by using special natures of 
low-dimensional systems.
%%%%%%%%%%%%%%%%%%%%%%%%%%%%%%%%%%%%%%
\subsection{Lieb-Mattis Theorem}
\label{s:LM}
Theorems discussed in the present and the next sections
state that the Hubbard model does not exhibit interesting long range 
order under some conditions.
The main purpose of studying an idealized model like 
the \Hub\ should be to 
show that some interesting physics {\em do} take place.
Results which say something {\em does not} take place may be regarded as 
less exciting.
But to have a definite knowledge that something does not happen under 
some conditions is very useful and important even if our final goal 
is to show something does happen.

The classical Lieb-Mattis theorem \cite{LiebMattis62} states (among other 
things) that one can never have ferromagnetism in one-dimensional \Hub\ 
with only nearest neighbor hoppings\footnote{
The present theorem appears in the Appendix of \cite{LiebMattis62}.
The main body of \cite{LiebMattis62} treats interacting electron 
systems in continuous spaces.
}.

\begin{theorem}[Lieb-Mattis theorem]
Consider a \Hub\ on a one-dimensional lattice $\La=\cbk{1,2,\ldots,N}$ with 
open boundary conditions.
We assume that the hopping matrix elements satisfy $|t_{x,y}|<\infty$ when 
$x=y$,  $0<|t_{x,y}|<\infty$ when $|x-y|=1$, and are vanishing 
otherwise.
We also assume $|U_x|<\infty$ for any  $x\in\La$. 
Then the quantity $\Emin(S)$ (see Definition \ref{Emin}) satisfies the 
inequality 
\begin{equation}
\Emin(S)<\Emin(S+1),
\label{E<E2}
\end{equation}
for any $S=0,1,\ldots,\Smax-1$ (or $S=1/2,3/2,\ldots,\Smax-1$).
\label{LiebMattis}
\end{theorem}

One of the most important consequences of the Lieb-Mattis theorem is that 
any one-dimensional \Hub\ in the above class has the total spin $\Stot=0$ 
(or $\Stot=1/2$) in its ground state.
One cannot conclude from this fact alone that the system exhibits
paramagnetism, but can conclude that there is no ferromagnetism.

Theorem \ref{LiebMattis} does not apply to models with periodic boundary 
conditions.
But it seems reasonable that boundary conditions do not 
change the essential physics provided that the system is sufficiently 
large.
If there exist hoppings to sites further than the nearest neighbor, on the 
other hand, the story is totally different.
We do not only find that the proof of Theorem \ref{LiebMattis} fails, 
but we also find essentially new physics.
See Section~\ref{secNew}.

Theorem \ref{LiebMattis} is proved by noting that, in a suitable 
basis, the Hamiltonian is written as a matrix whose nondiagonal elements 
are nonpositive, and by using the standard Perron-Frobenius argument.
Similar argument was used by Lieb and Mattis in their study of the 
Heisenberg quantum spin system \cite{LiebMattis62b}.
%%%%%%%%%%%%%%%%%%%%%%%%%%%%%%%%%%%%%%
\subsection{Decay of Correlations at Finite Temperatures}
\label{s:finite}
Among other rigorous results which show the absence of order is the 
extension by Ghosh \cite{Ghosh71} of the wellknown theorem of Mermin and 
Wagner.
Ghosh proved that the \Hub\ in one- or two-dimensions does not exhibit 
symmetry breaking related to magnetic long range order at any finite 
temperatures.
By using the same method, one can also prove the absence of 
superconducting  $U(1)$ symmetry breaking.

Koma and Tasaki proved essentially the same facts in terms of explicit 
upper bounds for various correlation functions \cite{92d}.
Among the results of \cite{92d} is the following.

\begin{theorem}[Koma-Tasaki bounds for correlations]
	\label{t:kt}
	Consider an arbitrary
	Hubbard model in one- or two-dimensions with finite 
	range hoppings.
	Then there are constants \( \alpha,\gamma \), and we have
\begin{equation}
	\abs{\bkt{
	\cd_{x,\up}\cd_{x,\dn}c_{y,\up}c_{y,\dn}+\mbox{\em H.c.}
	}_{\beta}} 
	\le
	\cases{ 
	|x-y|^{-\alpha f(\beta)}&\em in $d=2$;\cr
	\exp[-\gamma f(\beta)|x-y|]&\em in $d=1$,
	}
	\label{ktbound}
\end{equation}
	and
\begin{equation}
	\abs{\bkt{
	{\bf S}_{x}\cdot{\bf S}_{y}
	}_{\beta}} 
	\le
	\cases{ 
	|x-y|^{-\alpha f(\beta)}&\em in $d=2$;\cr
	\exp[-\gamma f(\beta)|x-y|]&\em in $d=1$,
	}
	\label{ktbound2}
\end{equation}
	for sufficiently large \( |x-y| \), where
	\( \bkt{\cdots}_{\beta} \) denotes the canonical average in 
	the thermodynamic limit at the inverse temperature $\beta$.
	Here \( f(\beta) \) is a decreasing function of \( \beta \) 
	and behaves as \( f(\beta) \approx 1/\beta\) for 
	\( \beta\gg\delta \) and 
	\( f(\beta) \approx (2/\delta)|\ln\beta| \) for 
	\( \beta\ll\delta \), where \( \delta \) is a constant.
\end{theorem}

The bounds (\ref{ktbound}) and (\ref{ktbound2})
establishes the widely accepted fact that there can be no
superconducting\footnote{
One can easily extend (\ref{ktbound}) to rule out the condensation of 
other types of electron pairs.
}
or magnetic long-range order
at finite temperatures in one- or two-dimensions.
The method employed in \cite{92d}, i.e., a combination of the 
McBryant-Spencer method and the quantum mechanical global $U(1)$ gauge 
invariance, is rather interesting.
It is amusing that only by using the $U(1)$ symmetry which exists in 
{\em any} quantum mechanical system, one gets upper bounds for 
correlations which are almost optimal at low temperatures
(especially in one-dimension).

%%%%%%%%%%%%%%%%%%%%%%%%%%%%%%%%%%%%%%
\subsection{Yamanaka-Oshikawa-Affleck Theorem}
\label{secYOA}
We discuss a recent important
theorem by Yamanaka, Oshikawa, and Affleck
\cite{YamanakaOshikawaAffleck97,OshikawaYamanaka97} about 
low-lying excitations in general electron systems on a 
one-dimensional lattice.
The theorem is an extension of Lieb-Schultz-Mattis theorem for quantum 
spin chains.
It can also be interpreted as a nonperturbative version of Luttinger's
``theorem'' restricted to one-dimension.

We consider the Hubbard model\footnote{
The theorem applies to a much larger class of lattice electron systems.
It is especially meaningful when applied to the Kondo lattice model.
} on the one-dimensional lattice
\( \La=\{1,2,\ldots,N\} \) with periodic boundary conditions.
The model is characterized by positive integers \( R \) and \( P \), 
which are the range of hopping and the period of the system, 
respectively.
We assume \( t_{x,y}=0 \) whenever \( |x-y|>R \), and
\( t_{x+P,y+P}=t_{x,y} \), \( U_{x+P}=U_{x} \) for any 
\( x,y\in\La \).
Under this general assumption, we have the following.

\begin{theorem}[Yamanaka-Oshikawa-Affleck theorem]
	\label{t:YOA}
	Consider the infinite volume limit
	\( N\to\infty \) with a fixed electron density
	\( \nu=\Ne/N \).
	If \( P\nu/2 \) is not an integer, then we have one of the following 
	two possibilities.
	\par\noindent
	i) There is a symmetry breaking, and the infinite volume ground 
	states are not unique.
	\par\noindent
	ii) There is a gapless excitation above the infinite volume ground 
	state.
\end{theorem}

In other words, the theorem rules out the third possibility;
\par\noindent
{\em
iii) The infinite volume ground state is unique, and there is a finite 
gap above it.
}

Note that iii) does take place if \( P\nu/2 \) is an integer and the 
system describes an innocent insulator\footnote{
Consider, for example, a two-band system with a band gap which has
\( P=2 \).
Then, in a free system, 
 \( \nu=1 \) (the half-filling) corresponds to an insulator with a 
charge gap.
}.
In a non-interacting system, it is evident that iii) is impossible 
when \( P\nu/2 \) is not an integer, since there is a partially 
filled band.
The above theorem guarantees that we cannot change the situation by 
introducing strong interaction.
This is of course far from obvious.

The proof of Theorem~\ref{t:YOA} is based on the following
explicit construction of a (trial) low-lying excitation\footnote{
An extra care is needed to discuss infinite volume limits
 \cite{OshikawaYamanaka97}.
}.
Consider a system on a periodic chain of length \( N \), and assume 
that the ground state \( \GS \) is unique.
We define the ``twist'' operator by
\( U=\exp[2\pi i\sum_{x=1}^{N}(x/N)n_{x,\up}] \)
which introduces a gradual twist in the \( U(1) \) phase of the up-spin 
electron field, and
consider the trial excited state
\( \Psi=U\GS \).
It is not hard to show that
\( \bkt{\Psi,H\Psi}-E_{\rm GS}=O(1/L) \).
Thus \( \Psi \) contains a low-lying excited state provided it is 
orthogonal to the ground state.
To see the orthogonality, let \( T_{P} \) be the translation by 
\( P \),
and assume that the ground state is chosen as \( T_{P}\GS=\GS \).
It is easily found that
\( T_{P}\Psi=e^{i\pi P\nu}\Psi \), and hence the trial state \( \Psi \)
is orthogonal to \( \GS \) whenever \( P\nu/2 \) is not an integer.

The above result also implies that, in the case ii), 
the gapless excitation
\( \Psi \) has a crystal 
momentum \( k_{\Psi}=\pi\nu \).
If we interpret this excitation as obtained from the 
ground state by moving an electron at a ``fermi point\footnote{
In a one-dimensional model, fermi surface (if any) becomes two ``fermi 
points.''}'' to the other fermi point, we find that 
the fermi momentum \( k_{\rm F} \) satisfies
\( k_{\Psi}=2k_{\rm F} \), and hence 
\( k_{\rm F}=\pi\nu/2 \).
This is nothing but the fermi momentum of the free system.
Therefore the fermi momentum is not ``renormalized'' by strong 
interaction among electrons.
As far as we know, this is the only general rigorous result which gives 
precise meaning to fermi momentum in truly interacting many electron 
systems.

%%%%%%%%%%%%%%%%%%%%%%%%%%%%%%%%%%%%%%
%%%%%%%%%%%%%%%%%%%%%%%%%%%%%%%%%%%%%%
\section{Half-Filled Systems}
\label{Sec1/2}
A system in which the electron number $\Ne$ is identical to the number 
of sites $|\La|$ is said to be half-filled, since the maximum possible 
value of  $\Ne$ is  $2|\La|$.
The system becomes half-filled if each atom provides one electron to the 
system.
Thus the half-filled models represent physically natural situations.
Half-filled models have nice properties\footnote{
Some half-filled models can be mapped onto a Hubbard model with attractive 
interaction via a partial hole-particle transformation.
This fact plays crucial role in the proof of Lieb's theorem.
} from mathematical point of view as well, and there are some very nice 
rigorous results.

%%%%%%%%%%%%%%%%%%%%%%%%%%%%%%%%%%%%%%
\subsection{Perturbation for $U\gg t$}
\label{secU>>t}
Let us first look at the ground states of the non-hopping model with 
$\Hhop=0$.
We here assume that $U_x>0$ for any  $x\in\La$. 
As we found in Section \ref{secnohop}, one can choose $X\cap Y=\emptyset$ 
in the state $\Psi_{X,Y}$ to get a ground state with  $E_{X,Y}=0$, 
provided that $\Ne\le|\La|$.
Since we have $\Ne=|\La|$, the assumption $X\cap Y=\emptyset$ 
automatically implies  $X\cup Y=\La$.
Therefore the ground state $\Psi_{X,Y}$  (\ref{PsiXY}) with $X\cap 
Y=\emptyset$ can be written also as
\begin{equation}
\Psi_{\tsigma}
=\rbk{\prod_{x\in\La}c^\dagger_{x,\sigma(x)}}
\vac,
\label{Psis}
\end{equation}
where $\tsigma=(\sigma(x))_{x\in\La}$ is a collection of spin indices 
$\sigma(x)=\up,\dn$.
By using the terminology of spin systems, one can call  $\tsigma$ a spin 
configuration.
One can say that the degeneracy of the ground states  (\ref{Psis}) 
precisely corresponds to all the possible spin configurations.

Let us take into account effects of nonvanishing $\Hhop$ by using a 
simple-minded perturbation theory. 
As the diagonal elements of $\Hhop$, i.e., 
$\sum_{x,\sigma}t_{x,x}\,\cxs\axs=\sum_xt_{x,x}(\nxu+\nxd)$, only shift 
the energy of the states $\Psi_{\tsigma}$ by a constant amount (independent 
of $\tsigma$), they can be 
omitted in the lowest order perturbation calculation.
Let us denote by $\tHhop=\sum_{x\ne y,\sigma}t_{x,y}\,\cxs\ays$ the 
off-diagonal part of $\Hhop$.
By operating $\tHhop$ once onto $\Psi_{\tsigma}$, an electron moves, and we 
get a state with one vacant site and one doubly occupied site.
The resulting state is not a ground state of $\Hint$.
We thus find that the lowest order contribution from this perturbation 
theory comes from the second order.

\begin{figure}
\centerline{\epsfig{file=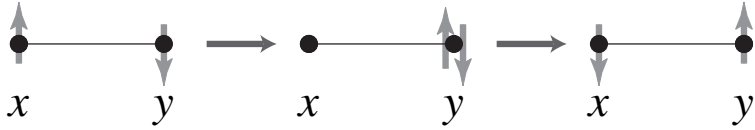,width=10cm}}
\caption[dummy]{
When electrons hop twice, spins on sites $x$ and $y$ may be exchanged.
This is the ultimate origin of antiferromagnetic nature in the half-filled 
\Hub.
}
\label{FIGpert}
\end{figure}

Figure \ref{FIGpert} shows a process that is taken into account in the 
second order perturbation theory.
The electron at site  $x$ hops to site  $y$ with the transition amplitude  
$t_{x,y}$, and generates a new state with extra energy $U_y$.
Then one of the two electrons at site  $y$ will hop back to site  $x$, and 
we recover one of the ground states.
In this process, spins at the sites  $x$ and $y$ may be exchanged as Figure 
\ref{FIGpert} shows.
The hopping between sites  $x$ and $y$ is inhibited by the Pauli principle 
if the electronic spins on these two sites are pointing in the same direction.
We find that this second order perturbation  process lowers the energy of 
states in which the spins at sites $x$ and $y$ are not pointing in the same 
direction (or more precisely, the states in which the total spin is 
vanishing).

Let us rederive this result in a more formal manner.
Let $P_0$ be the projection operator onto the subspace spanned by the 
states  $\Psi_{\tsigma}$ (\ref{Psis}) with all the possible $\tsigma$.
That the first order perturbation has no contribution can be read off from 
the fact that $P_0\tHhop P_0=0$.
To find out how the degeneracy in the (unperturbed) ground states
(\ref{Psis}) is lifted 
can be determined by the effective Hamiltonian
\begin{eqnarray}
H_{\rm eff}
&=&
-P_0\tHhop\frac{1}{\Hint}\tHhop P_0
\ret
&=&
P_0\cbk{\sum_{x,y\in\La}J_{x,y}\rbk{\Sop_x\cdot\Sop_y-\frac{1}{4}}}P_0.
\label{Heff}
\end{eqnarray}
Here the exchange interaction parameter is given by 
$J_{x,y}=\{(t_{x,y})^2/U_{x}\}+\{(t_{x,y})^2/U_{y}\}$.
Note that (\ref{Heff}) is nothing but the Hamiltonian of the $S=1/2$ 
antiferromagnetic quantum Heisenberg spin system.
This suggests that the low-energy behavior of the half-filled \Hub\ is 
well described by the Heisenberg antiferromagnets when $U_x$ are much 
larger than $t_{x,y}$.
%%%%%%%%%%%%%%%%%%%%%%%%%%%%%%%%%%%%%%
\subsection{Lieb's Theorem}
In 1989, Lieb proved an important and fundamental theorem for the 
half-filled \Hub.
The theorem provides, among other things, partial support to the 
conjecture about the similarity of the half-filled \Hub\ and the 
Heisenberg antiferromagnets.
Let us first introduce the notion of bipartiteness.

\begin{definition}[Bipartiteness]
Consider a Hubbard model (or other tight-binding electron model) on a 
lattice $\La$ with hopping matrix elements $\rbk{t_{x,y}}_{x,y\in\La}$.
The system is said to be bipartite if the lattice $\La$  can be decomposed 
into a disjoint union of two sublattices as $\La=A\cup B$ (with $A\cap 
B=\emptyset$), and it holds that $t_{x,y}=0$ whenever $x,y\in A$ or 
$x,y\in B$.
In other words, only hoppings between different sublattices are allowed.
\label{bipartite}
\end{definition}

Then Lieb's theorem \cite{Lieb89} for the repulsive \Hub\ is as follows.

\begin{theorem}[Lieb's theorem]
Consider a bipartite \Hub.
We assume $|\La|$ is even, and the whole $\La$ is connected\footnote{
More precisely, for any $x,y\in\La$, one can find a sequence of sites 
$x_0$, $x_1$, $\ldots$, $x_N$ with $x_0=x$, $x_N=y$, and 
$t_{x_i,x_{i+1}}\ne0$ for $i=0,1,\ldots,N-1$.
} through 
nonvanishing $t_{x,y}$.
We also assume $U_x=U>0$ for any  $x\in\La$.
Then the ground states of the model are nondegenerate apart from the 
trivial spin degeneracy\footnote{
In any quantum mechanical system with a rotation invariant Hamiltonian, an 
eigenstate of the Hamiltonian with the angular momentum  $J$ is always  
$(2J+1)$-fold degenerate. 
}, and have total spin $\Stot=\bigl||A|-|B|\bigr|/2$.
\label{Lieb}
\end{theorem}

The total spin $\Stot$ of the ground state determined in the theorem is 
exactly the same as that of the ground state(s) of the corresponding 
Heisenberg 
antiferromagnet on the same lattice.
In fact the conclusion of the theorem is quite similar to 
that of the Lieb-Mattis 
theorem \cite{LiebMattis62b} for Heisenberg antiferromagnets.
However the straightforward Perron-Frobenius argument used in the proof of 
the latter theorem does not apply to the \Hub\ except in one dimension.  
(See \ref{s:lowD}.)
This is not only a technical difficulty, but is a consequence of the 
important fact that quantum mechanical processes allowed in the \Hub\ are 
in general much richer and more complex than those in the Heisenberg model.
Lieb's proof is compactly presented in a Letter, but is deep and 
elegant.
The proof again makes use of a kind of Perron-Frobenius argument, but 
is based on an interesting technique called spin-space reflection 
positivity.

Lieb's theorem is valid for any value of Coulomb repulsion  $U$, only 
provided that it is positive.
It is quite likely that physical properties of the \Hub\ are drastically 
different in the weak coupling region with small  $U$ and in the strong 
coupling region with large  $U$.
It is very surprising and 
interesting that a single proof of Lieb's covers the whole range 
with  $U>0$ and clarifies the basic properties of the ground states.

It should be noted, however, that the knowledge of the total spin of 
the ground states in finite volume does not necessarily determine the 
properties of the ground states in the corresponding infinite system.
When two sublattices have the same number of sites as $|A|=|B|$, for 
example, one knows that finite volume ground state is unique and has 
$\Stot=0$.
Although one might well conclude that the system has no long range order 
in its ground states, this is not true.
It is certainly 
possible that infinite volume ground states exhibit long range 
order and symmetry breaking (such as antiferromagnetism of 
superconductivity) even when finite volume ground state is unique 
and symmetric.  
(See, for example, \cite{94b}.)
By knowing any finite volume ground state has  $\Stot=0$, however, 
one can rule out 
the possibility of ferromagnetism.

By using Lieb's results in \cite{Lieb89}, one gets some information about 
excited states.
For example, by combining Theorem 1 in \cite{Lieb89} with the method of 
\cite{LiebMattis62b}, one can easily prove the 
inequality\footnote{
I wish to thank Shun-Qing Shen and Elliott Lieb for discussions related to 
this corollary.
}
\begin{equation}
\Emin(S)<\Emin(S+1)
\label{new}
\end{equation}
for any $||A|-|B||/2\le S\le(|\La|/2)-1$.

Another theorem which suggests the similarity between the half-filled 
\Hub\ and the Heisenberg antiferromagnets is the following, proved 
by Shen, Qiu, and Tian \cite{ShenQiuTian94,Shen95} by 
extending Lieb's method.

\begin{theorem}[Explicit signs of correlations]
Assume the conditions for Theorem \ref{Lieb}.
If we denote by $\GS$ the ground state of the model, we have the 
inequalities
\begin{equation}
\bkt{\GS,\Sop_x\cdot\Sop_y\,\GS}
\cases{
>0& when $x,y\in A$, or $x,y\in B$;\cr
<0& when $x\in A,y\in B$, or $x\in B, y\in A$,\cr
}
\label{SQT}
\end{equation}
were $\bkt{\cdot,\cdot}$ denotes the quantum mechanical inner product.
\end{theorem}

We see that spins on different sublattices have negative correlations, 
indicating antiferromagnetic tendency.

The $S=1/2$ Heisenberg antiferromagnet on the cubic lattice, for 
example, is proved to exhibit an antiferromagnetic long range 
order at sufficiently low temperatures or in the ground states 
\cite{DysonLiebSimon78,KennedyLiebShastry88}.
It is likely that the same statements hold for the half-filled \Hub\ with 
sufficiently large  $U$. 
But, for the moment, there are no methods or ideas which are useful in 
proving the conjecture.
To extend the powerful (but not very natural) method of 
\cite{DysonLiebSimon78,KennedyLiebShastry88} based on the (spatial) 
reflection positivity seems hopeless.

In the Hubbard model and related models at half-filling, there have been 
proved several interesting general results.
Among the recent examples are the uniform density theorem 
\cite{LiebLossMcCann93}, the solution of the flux phase problem 
\cite{Lieb94,MacrisNachtergaele}, and the stability of the Peierls instability 
\cite{NachtergaeleLieb94}.
%%%%%%%%%%%%%%%%%%%%%%%%%%%%%%%%%%%%%%
\subsection{Lieb's Ferrimagnetism}
\label{secferi}
A very important corollary of Lieb's theorem \ref{Lieb} is that the 
half-filled Hubbard models on asymmetric bipartite lattices 
universally exhibit a 
kind of ferromagnetism (in the broad sense), or more precisely, 
ferrimagnetism \cite{Lieb89}. 

\begin{figure}
\centerline{\epsfig{file=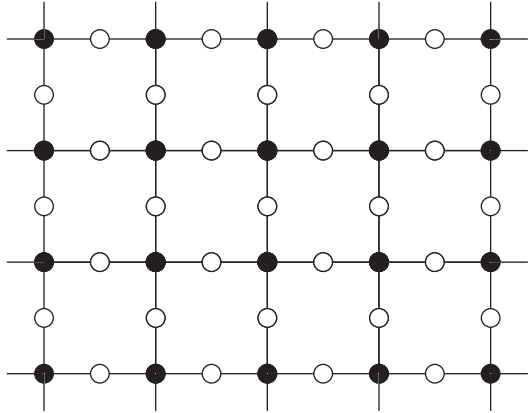,width=7cm}}
\caption[dummy]{
An example (the so called CuO lattice) of a bipartite lattice in which the 
number of sites in two sublattices are different.
Lieb's theorem implies that the half-filled \Hub\ defined on this lattice 
exhibits ferrimagnetism.
}
\label{FIGCuO}
\end{figure}

Take, for example, the so called CuO lattice in Figure \ref{FIGCuO}.
The lattice can be 
decomposed into two sublattices distinguished by black sites and white 
sites.
When the black sites form a square lattice with side  $L$, there are $L^2$ 
black sites and $2L^2$ white sites.
We define the \Hub\ on this lattice, and put nonvanishing hopping $t_{x,y}$ 
on each bond in the lattice, and put Coulomb interaction $U>0$ on each 
site.
Then Lieb's theorem implies that the ground state of this \Hub\ has total 
spin $\Stot=\bigl||A|-|B|\bigr|/2=L^2/2$.
Since the total spin magnetic moment of the system is proportional to the 
number of lattice sites $3L^2$, we conclude that the model exhibits 
ferromagnetism in the broad sense.

Of course the present ferromagnetism is not a saturated ferromagnetism in 
which all the spins in the system completely align with each other.
As the inequality (\ref{SQT}) suggests, spins on neighboring sites have 
tendency to point in the opposite direction. 
But the big difference in the numbers of sites in the sublattices cause 
the system to possess bulk magnetic moment.
Such a magnetic ordering is usually called ferrimagnetism\footnote{
It is also possible to consider order parameters to see the order is 
indeed ferrimagnetic \cite{ShenQiuTian94}.
}.

One can similarly construct Hubbard models
which exhibit ferrimagnetism on any 
bipartite lattice in which the difference in the number of sites in two 
sublattices is proportional to the system size.
The value of  $U>0$ is again arbitrary, so Lieb's ferrimagnetism covers 
surprisingly general class of models including weakly coupled ones as well 
as strongly coupled ones.

If one recalls the conclusion of Section \ref{secfree} that systems with 
$U=0$ exhibit paramagnetism, one might feel it somehow contradicting that 
the above ferrimagnetism appears for arbitrarily small  $U>0$.
This is one of the special features of Lieb's ferrimagnetism.
In the single-electron \Sch\ equation corresponding the \Hub\ on Figure 
\ref{FIGCuO}, for example, the eigenstates for the eigenvalue $\ep=0$ are 
$L^2$-fold degenerate.
(The eigenvalue $\ep=0$ is at the middle of the single-electron spectrum.)
Consequently the ground states of the half-filled ($\Ne=|\La|=3L^2$) 
system with $U=0$ are highly degenerate, and the total spin can take 
values in $\Stot=0,1,\ldots,L^2/2$.
The role of Coulomb interaction  $U$ is to lift this degeneracy, and 
select states with the largest magnetic moment as ground states.
%%%%%%%%%%%%%%%%%%%%%%%%%%%%%%%%%%%%%%
\subsection{Kubo-Kishi Bounds on  
Susceptibilities at Finite Temperatures}
\label{s:KK}
A theorem which can be regarded as a finite temperature version of Lieb's 
theorem was proved by Kubo and Kishi \cite{KuboKishi90}.
It deals with the charge susceptibility and the on-site pairing 
(superconducting)
susceptibility in a half-filled system at finite temperatures.

We define the thermodynamic function\footnote{
We have \( J=-pV=F-G \).
} 
\( J \) corresponding to the grand 
canonical ensemble by
\begin{eqnarray}
	&&
	J(\beta,\mu,(\gamma_{x})_{x\in\La},(\eta_{x})_{x\in\La})
	=
 	\ret
	&&
	=-\frac{1}{\beta}\log
	{\rm Tr}\exp\sqbk{
	-\beta\rbk{H-\mu\Neop
	-\sumtwo{x\in\La}{\sigma=\up,\dn}
	\gamma_{x}n_{x,\sigma}
	-\sum_{x\in\La}\eta_{x}(\cxu\cxd+\axd\axu)}
	},
	\label{GC}
\end{eqnarray}
where \( \beta \) and \( \mu \) are the inverse temperature and
the chemical potential, respectively,
and the trace is taken over the Hilbert spaces with all the possible electron 
numbers.
We added to the Hamiltonian two fictitious external fields 
\( (\gamma_{x})_{x\in\La} \) and 
\( (\eta_{x})_{x\in\La} \) to test for the possible charge ordering 
and superconducting ordering, respectively.

We define the charge susceptibility \( \chi^{\rm c} \) and
the on-site pairing susceptibility \( \chi^{\rm p} \) by
\begin{equation}
	\chi^{\rm c}_{\bf q}(\beta,\mu)
	=
	-\left.
	\frac{\partial}{\partial \tilde{\gamma}_{\bf q}}
	\frac{\partial}{\partial \tilde{\gamma}_{-\bf q}}
	J(\beta,\mu,(\gamma_{x}),(\eta_{x}))
	\right|_{(\gamma_{x})=(\eta_{x})=0}
	\ge0,
	\label{chic}
\end{equation}
and
\begin{equation}
	\chi^{\rm p}_{\bf q}(\beta,\mu)
	=
	-\left.
	\frac{\partial}{\partial \tilde{\eta}_{\bf q}}
	\frac{\partial}{\partial \tilde{\eta}_{-\bf q}}
	J(\beta,\mu,(\gamma_{x}),(\eta_{x}))
	\right|_{(\gamma_{x})=(\eta_{x})=0}
	\ge0.
	\label{chip}
\end{equation}
The Fourier transformation of the external fields are
\begin{equation}
	\tilde{\gamma}_{\bf q}=
	|\La|^{-1/2}\sum_{x\in\La}\gamma_{x}\,e^{i{\bf q}\cdot x},
	\quad
	\tilde{\eta}_{\bf q}=
	|\La|^{-1/2}\sum_{x\in\La}\eta_{x}\,e^{i{\bf q}\cdot x},
	\label{Fouries}
\end{equation}
where \( \bf q \) is a wave vector corresponding to the lattice
\( \La \) (which we assume to have a periodic structure).

Then the Kubo-Kishi theorem states that

\begin{theorem}
	\label{t:KK}
	Consider any bipartite (see Definition~\ref{bipartite}) Hubbard 
	model with \( U_{x}=U>0 \) for any \( x\in\La \).
	Then for any \( \beta>0 \) and for any wave vector \( \bf q \),
	we have
	\begin{equation}
		\chi^{\rm c}_{\bf q}(\beta,U/2)\le\frac{1}{U},
		\quad
		\mbox{\em and}
		\quad
		\chi^{\rm p}_{\bf q}(\beta,U/2)\le\frac{2}{U}.
		\label{KKbounds}
	\end{equation}
\end{theorem}

Note that the choice \( \mu=U/2 \) corresponds to the half-filling.
The theorem states that the charge and the on-site paring 
susceptibilities for any wave vector \( \bf q \) are finite in a 
half-filled model at finite temperatures.
This means that the model does not exhibit any CDW ordering or 
superconducting ordering.

%%%%%%%%%%%%%%%%%%%%%%%%%%%%%%%%%%%%%%
%%%%%%%%%%%%%%%%%%%%%%%%%%%%%%%%%%%%%%
\section{Ferromagnetism}
Ferromagnetism, where almost all of the spins in the system align in the 
same direction, is a remarkable phenomenon.
Standard theories about the origin of ferromagnetism have been the 
Heisenberg's exchange interaction picture, and the Stoner criterion 
derived from the Hartree-Fock approximation for band electrons.
But there have been serious doubts if these theories really explain the 
appearance of ferromagnetism in a system of  electrons interacting via 
spin-independent Coulomb interaction.
One of the motivations to study the \Hub\ was to understand the origin of 
ferromagnetism in an idealized situation.

As we have seen in the previous section, half-filled models have tendency 
towards antiferromagnetism.
In this section we shall concentrate on systems in which the electron 
numbers deviate from half-filling.
%%%%%%%%%%%%%%%%%%%%%%%%%%%%%%%%%%%%%%
\subsection{Instability of Ferromagnetism}
\label{s:noferro}
To see that ferromagnetism is indeed a delicate phenomenon, we  
discuss two elementary
results which show that the Hubbard model with certain 
conditions does {\em not} exhibit ferromagnetism\footnote{
Detailed proofs of the results in the present section can be found in
\cite{97c}.
}.

The following theorem states that 
there can be no saturated ferromagnetism if 
the Coulomb interaction \( U \) 
is too small in a system with a ``healthy'' single-electron spectrum.

\begin{theorem}[Impossibility of ferromagnetism for small \( U \)]
\label{t:var1}
Let \( \{\ep_{j}\}_{j=1,\ldots,N} \) denote the single-electron 
energy eigenvalues with \( \ep_{j}\le\ep_{j+1} \) as in 
Section~\ref{secfree}.
If
\( 0\le U<\ep_{\Ne}-\ep_{1} \),
we have\footnote{
Note that the fermi energy \( \ep_{\Ne}-\ep_{1} \) is an intensive 
quantity.
}
\begin{equation}
\Emin(\Smax-1)<\Emin(\Smax).
\label{E<E}
\end{equation}
Thus the ground state of the model does not have
\( \Stot=\Smax \).
\end{theorem}

When the density of electrons is very low, 
it is expected that the chance of electrons to collide with each other 
becomes very small.
It is likely that the model is close to an ideal gas
no matter how strong the interaction is, and there is no 
ferromagnetism.

This naive guess is easily justified for ``healthy'' models in 
dimensions three (or higher).
The dimensionality of the lattice is taken into account by assuming 
that there are positive constants \( c \), \( \nu_{0} \), and 
\( d \), and the single electron energy levels satisfy
\begin{equation}
	\ep_{n}-\ep_{1}\ge c\rbk{\frac{n-1}{\Ns}}^{2/d},
	\label{healthy}
\end{equation}
for any \( n \) such that 
\( n/\Ns\le\nu_{0} \).
Note that the right-hand side represents the \( n \) dependence of 
energy levels in an usual \( d \)-dimensional quantum mechanical 
system.
Then we have the following theorem due to
Pieri, Daul, Baeriswyl, Dzierzawa, and Fazekas \cite{Pieri97}.

\begin{theorem}[Impossibility of ferromagnetism at low densities]
	\label{t:var2}
	Suppose we have \( \Hhop \) satisfying (\ref{healthy}) with positive 
	\( c \), \( \nu_{0} \), and \( d>2 \).
	Then there exists a constant \( \nu_{1}>0 \), and the same conclusion 
	as Theorem~\ref{t:var1} holds for any \( U\ge0 \)
	if \( \Ne/\Ns\le\nu_{1} \) holds.
\end{theorem}

That we have a restriction on dimensionality in Theorem~\ref{t:var2} 
is not merely technical.
In a one-dimensional system, moving electrons must eventually collide 
with each other for an obvious geometric reason.
Thus a one-dimensional model cannot be regarded as close to ideal no 
matter how low the electron density is.
We do not know whether the inapplicability of the theorem to
\( d=2 \) systems is physically meaningful or not.

%%%%%%%%%%%%%%%%%%%%%%%%%%%%%%%%%%%%%%
\subsection{Toy Model with Two Electrons}
\label{sectoy}
As a starting point of our study of ferromagnetism, we consider a toy 
model with two electrons on a small lattice.
Interestingly enough, some essential features of ferromagnetism 
found in many-electron systems (that we will discuss later in this 
section) are already present in the toy model.

\begin{figure}
\centerline{\epsfig{file=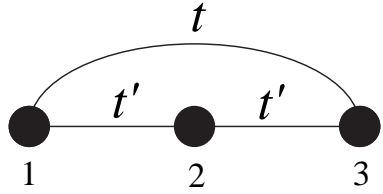,width=5cm}}
\caption[dummy]{
The lattice and the hopping of the toy model.
By considering the system with two electrons on this lattice, we can 
observe some very important aspects of ferromagnetism in the \Hub.
}
\label{FIGtoymodel}
\end{figure}

The smallest possible model in which we can discuss electron interaction 
and which is away from half-filling is that with two electrons on a 
lattice with three sites.
Consider the lattice $\La=\cbk{1,2,3}$, and put one electron with 
$\sigma=\up$ and one with $\sigma=\dn$.
The hopping matrix is defined by $t_{1,2}=t_{2,3}=t'$, and $t_{1,3}=t$.
Note that there are two kinds of hoppings  $t$ and  $t'$.  
Since the sign of $t'$ can be changed by the gauge transformation 
$c_{2,\sigma}\to-c_{2,\sigma}$, we shall fix $t'>0$.
Figure \ref{FIGtoymodel} shows the lattice and the hopping.
For simplicity, we assume there is only one kind of interaction, and set 
$U_{1}=U_{2}=U_{3}=U\ge0$.
We have $\Smax=1$ because $\Ne=2$.
Therefore we can say that there appears saturated ferromagnetism if the 
ground state has $\Stot=1$, i.e., if it is a part of a spin-triplet.

\begin{figure}
\centerline{\epsfig{file=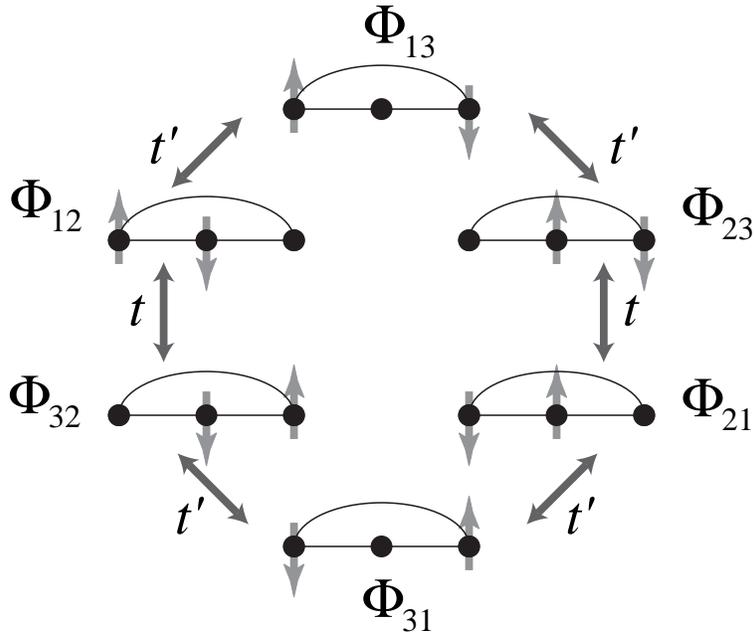,width=10cm}}
\caption[dummy]{
Allowed states and transition amplitudes in the toy model with $U=\infty$.
The total spin of the ground states can be easily read off from this 
diagram.
}
\label{FIGtoy}
\end{figure}

Let us take the limit $U\to\infty$, in which the effect of interaction 
becomes most drastic, and consider only those states with finite energies.
This is equivalent to consider only states in which two electrons never 
occupy a same site.
There are six states which satisfy the constraint, and they can be written 
as $\Phi_{x,y}=c^{\dagger}_{x,\up}c^{\dagger}_{y,\dn}\vac$ where 
$x,y=1,2,3$, and $x\ne y$.
Transition amplitudes between these states are shown in Figure \ref{FIGtoy}.
We find that the problem is equivalent to that of a quantum mechanical 
particle hopping around on a ring consisting of six sites.
The basic structure of the ground state can be determined by the standard 
Perron-Frobenius sign convention\footnote{
If the transition amplitude between two states is negative ({\em resp.}, 
positive), one superposes the two states with the same ({\em resp.}, 
opposite) signs.
}.
The ground state for $t<0$ is written as
\begin{equation}
\GS^{(t<0)}=
\Phi_{1,2}+\Phi_{3,2}-\alpha(t,t')\Phi_{3,1}+\Phi_{2,1}+\Phi_{2,3}
-\alpha(t,t')\Phi_{1,3},
\label{GSt<0}
\end{equation}
and that for $t>0$ as
\begin{equation}
\GS^{(t>0)}=\Phi_{1,2}-\Phi_{3,2}+\beta(t,t')\Phi_{3,1}
-\Phi_{2,1}+\Phi_{2,3}-\beta(t,t')\Phi_{1,3},
\label{GSt>0}
\end{equation}
where $\alpha(t,t')$ and $\beta(t,t')$ are positive functions of $t$ and 
$t'$.

To find the total spin of these states, it suffices to concentrate on two 
lattice sites, say sites 1 and 2, and note that 
$\GS^{(t<0)}=\Phi_{1,2}+\Phi_{2,1}+\cdots$, and 
$\GS^{(t>0)}=\Phi_{1,2}-\Phi_{2,1}+\cdots$.
It immediately follows that\footnote{
A quick way to find the total spin of the state $\Phi_{1,2}+\Phi_{2,1}$ is 
to rewrite the state in the ``spin language'' as 
$\Phi_{1,2}+\Phi_{2,1}
=c^{\dagger}_{1,\up}c^{\dagger}_{2,\dn}\vac
+c^{\dagger}_{2,\up}c^{\dagger}_{1,\dn}\vac 
=c^{\dagger}_{1,\up}c^{\dagger}_{2,\dn}\vac
-c^{\dagger}_{1,\dn}c^{\dagger}_{2,\up}\vac 
=\ket{\up}_{1}\ket{\dn}_{2}-\ket{\dn}_{1}\ket{\up}_{2}$,
and use the standard knowledge about the addition of angular momenta.
One can easily convince oneself that the state has $\Stot=0$.
} $\GS^{(t<0)}$ has $\Stot=0$, and $\GS^{(t>0)}$ has $\Stot=1$.
A ferromagnetic coupling is generated when  $t>0$!

Let us look at the mechanism which generates the ferromagnetism.
The states $\Phi_{1,2}$ and $\Phi_{2,1}$ can be found in the upper left 
and and the lower right, respectively, in the diagram of Figure 
\ref{FIGtoy}.
By starting from $\Phi_{1,2}$ and following the possible transitions, one 
reaches the state $\Phi_{2,1}$.
In other words, electrons hop around in the lattice, and the spins on 
sites 1 and 2 are ``exchanged.''
When $t>0$, the quantum mechanical amplitude associated with the exchange 
process generates the superposition of the two states which precisely 
yields ferromagnetism.

\begin{figure}
\centerline{\epsfig{file=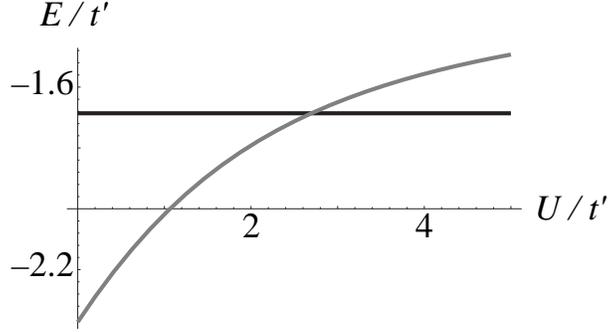,width=8cm}}
\caption[dummy]{
The $U$ dependence of $\Emin(0)$ (gray curve) and $\Emin(1)$ (black line) 
in the toy model with $t=t'/2$.
We have ferromagnetism in the sense that $\Emin(0)>\Emin(1)$ when $U$ is 
sufficiently large.
We find that ferromagnetism is a ``nonperturbative'' phenomenon.
}
\label{FIGgraph1}
\end{figure}

Let us briefly look at the cases with finite  $U$.
In Figure  \ref{FIGgraph1}, we plotted $\Emin(0)$ and $\Emin(1)$ for the 
toy model with $t=t'/2$ as functions of  $U$. 
(See Definition \ref{Emin}.)
As is suggested by the result in the $U\to\infty$ limit, we have 
ferromagnetism in the sense that $\Emin(0)>\Emin(1)$ when $U$ is 
sufficiently large.
A level crossing takes place at finite $U$, and the system is no longer 
ferromagnetic for small $U$.
Even in the simplest toy model, ferromagnetism is a ``nonperturbative'' 
phenomenon which takes place only when $U$ is sufficiently large.

\begin{figure}
\centerline{\epsfig{file=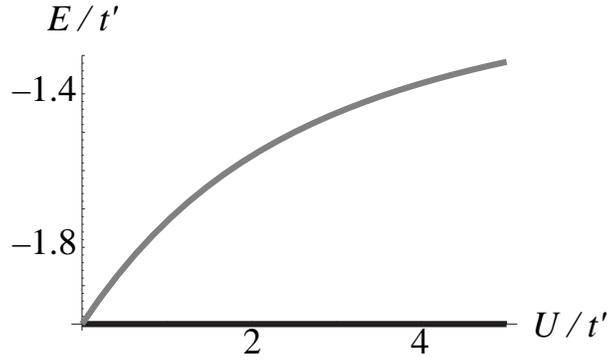,width=8cm}}
\caption[dummy]{
The $U$ dependence of $\Emin(0)$ (gray curve) and $\Emin(1)$ (black line) 
in the toy model with $t=t'$.
Only for this special parameter, we have ferromagnetism $\Emin(0)>\Emin(1)$ 
for any value of $U>0$.
One can regard this case as the simplest example of the flat-band 
ferromagnetism that we will discuss in Section \ref{secflat}.
}
\label{FIGgraph2}
\end{figure}

The only exception is the case with $t=t'$.
See Figure \ref{FIGgraph2}.
For this parameter value, the ground states are degenerate in spin when 
$U=0$.
Ferromagnetic state is the only ground state for $U>0$.

  From Figures \ref{FIGgraph1} and \ref{FIGgraph2}, we find that the energy 
$\Emin(1)$ of ferromagnetic states is independent of $U$.
As we see in the following, this is a general property of ferromagnetic 
eigenstates in the \Hub.
An arbitrary state $\Psi$ which has total spin $\Stot=\Smax$ can be 
written as a superposition of states which are obtained by rotating the 
state $\widetilde{\Psi}$ which consists only of $\up$ spin electrons.
If one operates the interaction Hamiltonian $\Hint$ (\ref{Hint}) onto the 
state $\widetilde{\Psi}$, one has $\Hint\widetilde{\Psi}=0$ because 
$n_{x,\dn}\widetilde{\Psi}=0$.
Since the interaction Hamiltonian (\ref{Hint}) is invariant under 
rotation in spin space, we have shown that $\Hint\Psi=0$.
One might say that states with saturated magnetization do not feel Hubbard 
type interaction at all.
This is one of convenient 
(but ``oversimplified'') features in discussing saturated ferromagnetism 
in the \Hub.
%%%%%%%%%%%%%%%%%%%%%%%%%%%%%%%%%%%%%%
\subsection{Nagaoka's Ferromagnetism}
The transitions between the states in Figure \ref{FIGtoy} are generated by 
hoppings of electrons.
One can also regard that the transitions are caused by hoppings of a 
single hole\footnote{
This is different from the notion of hole in the usual band theory.
}, which is the site without electrons.
At least in the limit $U\to\infty$, one can say that the origin of 
ferromagnetism in the toy model is the motion of a single hole, which mix 
up various spin configurations with proper signs.

\begin{figure}
\centerline{\epsfig{file=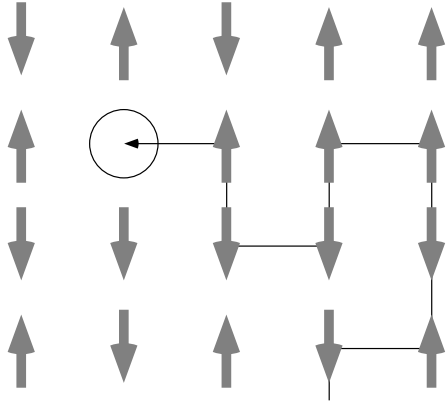,width=6cm}}
\caption[dummy]{
Schematic picture about the origin of Nagaoka's ferromagnetism.
When the hole hops around the lattice, the spin configuration is 
changed.
For a model with \( t_{x,y}\ge0 \), the hole motion produces a precise 
linear combination of various spin configurations which leads to 
ferromagnetic state.
}
\label{f:nagaoka}
\end{figure}

As Nagaoka \cite{Nagaoka66} demonstrated rigorously, 
there is a class of many electron 
models in which saturated ferromagnetism is generated by exactly the same 
mechanism.
See Figure~\ref{f:nagaoka}.
Nagaoka's theorem (in the extended form of \cite{89c} whose complete 
proof can be found in \cite{97c}) is as follows.

\begin{theorem}[Nagaoka's ferromagnetism]
Take an arbitrary finite lattice $\La$, and assume that\footnote{
As we noted in Section \ref{s:def}, this sign of $t_{x,y}$ is opposite 
from the ``standard'' choice.
In bipartite systems (such as those on the square lattice or the cubic 
lattice with nearest neighbor hoppings), one can change the sign of 
$t_{x,y}$ by a gauge transformation.
} $t_{x,y}\ge0$ for any $x\ne y$, and $U_{x}=\infty$ for any  $x\in\La$. 
We fix the electron number as $\Ne=|\La|-1$.
Then among the ground states of the model, there exist states with total 
spin $\Stot=\Smax(=\Ne/2)$.
If the system further satisfies the connectivity condition, then the ground 
states have $\Stot=\Smax(=\Ne/2)$ and are nondegenerate apart from the 
trivial spin degeneracy.
\label{Nagaoka}
\end{theorem}

The connectivity condition is a simple condition which holds on most of 
the lattices in two or higher dimensions, including the square lattice, 
the triangular lattice, or the cubic lattice.
To be precise the condition requires that ``by starting from any electron 
configuration on the lattice and by moving around the hole along 
nonvanishing $t_{x,y}$, one can get any other electron configuration.''

Thouless also reached a similar conclusion \cite{Thouless65}, but 
Nagaoka's treatment covers a larger class of models including 
non-bipartite systems.
The proof of Nagaoka's theorem (especially the recent proof in \cite{89c}) 
is surprisingly simple.
It essentially uses the Perron-Frobenius argument exactly the same as that 
we used in Section \ref{sectoy} to determine the total spin of the ground 
state of the toy model.

The requirements that $U$ should be infinitely large and there should be 
exactly one hole are indeed rather pathological.
Nevertheless, the theorem is very important since it showed for the first 
time in a rigorous manner that quantum mechanical motion of electrons and 
strong Coulomb repulsion can generate ferromagnetism.
The conclusion that the system which has one less electron than the 
half-filled model exhibits ferromagnetism is indeed surprising.
This is a very nice example which demonstrates that 
strongly interacting electron systems  
produce very rich physics.

It is desirable to extend Nagaoka's ferromagnetism to systems with  a 
finite  $U$ and with a finite density of holes.
Although more than thirty years have passed since Nagaoka's and Thouless's 
papers, it is still not known if such extensions are possible.
There are, however, considerable amount of rigorous works which 
establish that saturated ferromagnetism does not take place in 
certain situations.
See, for example, 
\cite{DoucotWen89,Shastry90,Toth91,Suto91b,HanischMullerHartmann93,%
HanischUhrigMuellerHartmann97}.

%%%%%%%%%%%%%%%%%%%%%%%%%%%%%%%%%%%%%%
\subsection{Mielke's Ferromagnetism and Flat-Band Ferromagnetism}
\label{secflat}
Let us once again look at the toy model of section \ref{sectoy}.
As is shown in Figure \ref{FIGgraph2}, the ground state of the model 
exhibits ferromagnetism for any $U>0$ for the special choice of the 
parameters $t=t'>0$.
For this choice of parameters, the energy eigenvalues of the 
corresponding \Sch\ equation are $\ep_{1}=\ep_{2}=-t'$, and $\ep_{3}=2t'$.
The single-electron ground states are doubly degenerate.
As a consequence, the ground states of the two electron system with  $U=0$ 
are also degenerate, and can have $\Stot=0$ or $\Stot=1$.
The degeneracy is lifted for $U>0$, and the ferromagnetic state is 
``selected'' as the true and unique ground state.
It is crucial here that the dimension of the degeneracy in the 
single-electron ground states (which is two) is the same as the electron 
number  $\Ne=2$.

\begin{figure}
\centerline{\epsfig{file=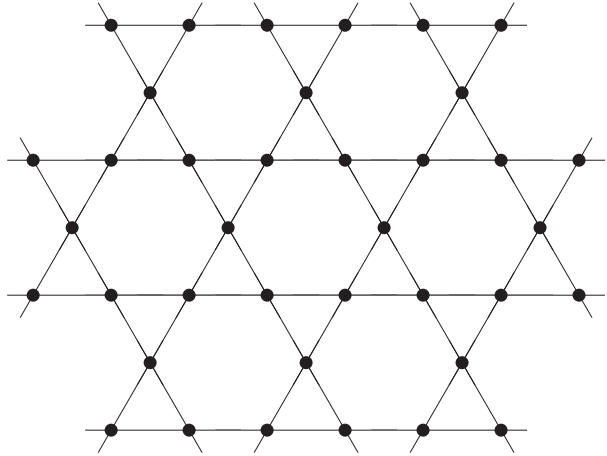,width=8cm}}
\caption[dummy]{
The \Hub\ on the kagom\'{e} lattice is a typical example which exhibits 
flat-band ferromagnetism.
``Kagom\'{e}'' is a Japanese word for a pattern of woven bamboo 
of baskets.
}
\label{FIGkagome}
\end{figure}

Mielke \cite{Mielke91b} showed that there is a class of Hubbard models
with many 
electrons which show saturated ferromagnetism through a somewhat similar 
mechanism.
Take, for example, the kagom\'{e} lattice of Figure \ref{FIGkagome}, and 
define a \Hub\ on it by setting $t_{x,y}=t>0$ for neighboring sites  
$x,y$,  $t_{x,y}=0$ for other situations, and $U_{x}=U\ge0$ for any  
$x\in\La$.
It is worth mentioning that the kagom\'{e} lattice of Figure 
\ref{FIGkagome} can be regarded as constructed by putting together many 
copies of the lattice used in the toy model (Figure \ref{FIGtoymodel}).
The energy eigenvalues of the corresponding \Sch\ equation can be shown 
to satisfy $\ep_{1}=\ep_{2}=\cdots=\ep_{M}=-2t$, and $\ep_j>-2t$ for $j>M$.
Here the dimension  $M$ of the degeneracy of the single-electron ground 
states is given by $M=(|\La|/3)+1$, and is proportional to the lattice 
size.

We shall fix the electron number as $\Ne=M$, the same as the dimension of 
the degeneracy.

Let us consider the case with $U=0$ first.
Let $A$ and $B$ be arbitrary subsets of $\cbk{1,2,\ldots,\Ne}$ which 
satisfy $|A|+|B|=\Ne$, and consider the state $\Phi_{A,B}$ (\ref{PhiAB}) 
obtained by creating the corresponding single-electron eigenstates.
In the present model on the kagom\'{e} lattice, the fermion operator 
$a^{\dagger}_{j,\sigma}$ (\ref{a}) creates one of the single-electron 
ground states with the energy  $\ep=-2t$.
This means that we have $\Hhop\Phi_{A,B}=-2t\Ne\Phi_{A,B}$ 
for arbitrary choice of  $A$ 
and  $B$, and hence $\Phi_{A,B}$ is a ground state of $H=\Hhop$.
We find that the ground states are highly degenerate, and can have 
$\Stot=0,1,\ldots,\Smax=M/2$ (or $\Stot=1/2,\ldots,\Smax$).

What is the effect of nonvanishing Coulomb interaction  $U$ in such a 
situation?
Let us denote by $\Phi_{\up}$ the state obtained by setting 
$A=\cbk{1,2,\ldots,\Ne}$ and $B=\emptyset$ in $\Phi_{A,B}$.
Of course $\Phi_{\up}$ is one of the ground states of $\Hhop$.
As we discussed at the end of Section \ref{sectoy}, the state $\Phi_{\up}$ 
which consists only of $\up$ spin electrons ``does not feel'' Hubbard type 
Coulomb interaction.
This means that we have $\Hint\Phi_{\up}=0$.
Since  $0$ is the minimum possible eigenvalue of $\Hint$, we find that the 
state $\Phi_{\up}$ is a ground state of the total 
Hamiltonian$H=\Hhop+\Hint$ for any $U>0$.

These are all simple observations.
A really interesting problem is whether there can be ground 
states other than $\Phi_{\up}$ when $U>0$.
The following theorem due to Mielke shows that the ferromagnetic state is 
indeed ``selected'' as the true ground state exactly as in the toy model.

\begin{theorem}[Mielke's flat-band ferromagnetism]
Consider the \Hub\ on the kagom\'{e} lattice described above.
For any $U>0$, the ground states have $\Stot=\Smax(=M/2)$ and are 
nondegenerate apart from the trivial spin degeneracy.
\label{Mielke}
\end{theorem}

Mielke \cite{Mielke92} also extended his results to the situation where the 
electron density\footnote{
There is a minor error in the derivation of critical electron density in 
Mielke's paper.
One should modify this part by using the method of \cite{93d}.
} $\Ne/|\La|$ is less than  $1/3$ but close to  $1/3$.

In Mielke's work, it was proved for the first time that 
the \Hub\ with finite  $U$ can 
exhibit saturated ferromagnetism.
The model is very simple, and the result is very important.
As far as the author knows, there had been no discussions about 
the possibility of ferromagnetism in the \Hub\ on the kagom\'{e} lattice.
Mielke's work is not only mathematically rigorous, but important from 
physicists' point of view as it opened a new way of approaching itinerant 
electron ferromagnetism.

Mielke's proof of his main theorem is an elegant induction which makes use 
of a graph theoretic language.
The proof is not at all trivial since the problem is intrinsically a 
many-body one.
However, there is a very special feature of the model that any ground 
state of the total Hamiltonian $H=\Hhop+\Hint$ is at the same time a ground 
state of each of $\Hhop$ and $\Hint$.
Because of this property, one does not have to face the very difficult 
problem in many-body problems called the ``competition between $\Hhop$ 
and $\Hint$.''
That one has ferromagnetism in this model for any $U(>0)$ is closely 
related to this fact.

Mielke's theorem applies not only to the \Hub\ on the kagom\'{e} lattice 
but to those on a wide class of lattices called line graphs.
In all of these models, the ground states in the corresponding 
single-electron \Sch\ equation are highly degenerate.
There have been constructed \cite{92e,93d} other examples of Hubbard models
 in 
which the corresponding single-electron ground states are highly 
degenerate, and exhibit saturated ferromagnetism for any  $U>0$.
Ferromagnetism in the examples of Mielke and in \cite{92e,93d} are now 
called flat-band ferromagnetism\footnote{
  From the view point of band structure in the single-electron problem, the 
bulk degeneracy in the single-electron ground states correspond to the 
lowest band being completely dispersionless (or flat).
}.
Mielke \cite{Mielke93} obtained a necessary and sufficient condition for a 
\Hub\ with highly degenerate single-electron ground states to exhibit 
saturated ferromagnetism.
It is interesting that Lieb's ferrimagnetism discussed in Section 
\ref{secferi} resembles flat-band ferromagnetism in that the 
corresponding single-electron spectrum has a bulk degeneracy.

It is needless to say that the models in which single-electron ground 
states are highly degenerate are rather singular.
By adding a generic small perturbation to the hopping Hamiltonian, the 
degeneracy is lifted in general, and one gets a nearly flat lowest band 
rather than a completely flat one.
A very interesting and important problem is whether ferromagnetism remains 
stable after such a perturbation is added.
Of course one does not have ferromagnetism for small enough  $U$ if the 
bulk degeneracy in the single-electron ground states is lifted.
What one expects is that ferromagnetism remains stable when  $U$ is 
sufficiently large.
(Recall that in the toy model of Section \ref{sectoy}, we had 
ferromagnetism for all $U>0$ only for the special choice of parameters 
$t=t'$.)
There are some indications (from numerical or variational calculations) 
that ferromagnetism is stable under perturbation.
As for rigorous results, stability of ferromagnetism under single-spin 
flip is proved in \cite{94c,95b} for the model obtained by adding 
arbitrary small perturbation to the \Hub\ of \cite{92e,93d}.
As for a special class of perturbations, 
the problem of stability of ferromagnetism 
is completely solved as we shall see in the next section.

%%%%%%%%%%%%%%%%%%%%%%%%%%%%%%%%%%%%%%
\subsection{Ferromagnetism in a Non-Singular \Hub}
\label{secNew}
We have seen two  theorems which show that certain Hubbard models exhibit 
saturated ferromagnetism.
In Nagaoka's theorem, it is assumed that the system has exactly one hole, 
and has infinitely large Coulomb interaction.
In Mielke's theorem and other flat-band ferromagnetism, it is essential 
that the single-electron ground states have a bulk degeneracy.
Is it possible to prove the existence of saturated ferromagnetism in a 
non-singular \Hub\ which have finite $U$ and in which the single-electron 
spectrum is not singular?
Recently such examples were constructed \cite{95c}.

\begin{figure}
\centerline{\epsfig{file=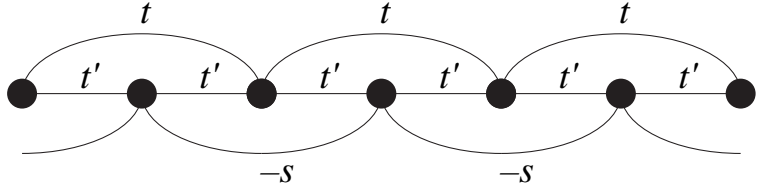,width=10cm}}
\caption[dummy]{
An example of non-singular \Hub\ which exhibits saturated ferromagnetism 
\cite{95c}.
If we look at three adjacent sites, the lattice structure and the hopping 
resemble that of the toy model of Figure \ref{FIGtoymodel}.
}
\label{FIGHal}
\end{figure}

For simplicity, we concentrate on the simplest models in one 
dimension\footnote{
There are models in higher dimensions \cite{97e}.
In the original paper \cite{95c}, the model contains an additional 
parameter $\lambda>0$.
Here we have set $\lambda=\sqrt{2}$ to simplify the discussion.
The proof of the main theorem in \cite{95c} is considerably improved 
in \cite{97e}.
The condition \( \lambda\ge\lambda_{\rm c} \) in \cite{95c} is 
replaced by \( \lambda>0 \).
}.
Take the one dimensional lattice $\La=\cbk{1,2,\ldots,N}$ with  $N$ sites 
(where  $N$ is an even integer), and impose a periodic boundary 
condition by identifying the site  $N+1$ with the site  $1$.
The hopping matrix is defined by setting $t_{x,x+1}=t_{x+1,x}=t'$ for any 
$x\in\La$, $t_{x,x+2}=t_{x+2,x}=t$ for even  $x$, $t_{x,x+2}=t_{x+2,x}=-s$ 
for odd  $x$, and  $t_{x,y}=0$ otherwise.
Here $t>0$ and $s>0$ are independent parameters, but the parameter $t'$ is 
determined as $t'=\sqrt{2}(t+s)$.
As can be seen from Figure \ref{FIGHal}, the model\footnote{
Solvable Hubbard models with $U=\infty$ which have similar 
structure as the present models were found by Brandt and Giesekus 
\cite{BrandtGiesekus92}, and were extended in \cite{M,S,T1,T2}.
The conjectured uniqueness of the ground state was proved in 
\cite{BL,T2}.
The ground state correlation functions in one dimensional models were 
calculated exactly in \cite{BL,Y}, and insulating behavior was found.
(\cite{T1} contains an error which is corrected in the footnote~6 of 
\cite{T2}.
Although I discussed the possibility of superconductivity in these model 
in \cite{T1}, 
I am not very optimistic about this conjecture at present.)
} has two kinds of next 
nearest neighbor hoppings $t$ and $-s$, as well as the nearest neighbor 
hopping $t'$.
If we look at an odd site and the two neighboring even sites, the model is 
exactly the same as the toy model we treated in Section \ref{sectoy}.
Roughly speaking, this resemblance is the basic origin of ferromagnetism in 
the present model.
We also note that because there are next nearest neighbor hoppings, 
Lieb-Mattis theorem (Theorem \ref{LiebMattis}) does not apply to the 
present model.

The single electron energy eigenvalue in this model can be expressed by 
using the wave number $k=2\pi n/N$ 
($n=0,\pm1,\ldots,\pm\cbk{(N/4)-1},N/4$) as $\ep_{1}(k)=-2t-2s(1+\cos 2k)$ 
and $\ep_{2}(k)=2s+2t(1+\cos 2k)$.
There are two bands, and each of them has healthy dispersion.

As for the Coulomb interaction, we set $U_{x}=U>0$ for any  $x\in\La$.
We fix the electron number as $\Ne=N/2$.
In terms of filling factor, this corresponds to the quarter 
filling.
The maximum possible value of total spin is $\Smax=N/4$.

Unlike in flat-band ferromagnetism, there is no saturated ferromagnetism 
when  $U$ is sufficiently small.
Theorem~\ref{t:var1} ensures that the ground state have 
\( \Stot<\Smax \) if \( U<4s \).
If the present system were to show saturated ferromagnetism, it should be 
in the ``nonperturbative'' region with sufficiently large $U$.
The following theorem of \cite{95c} provides such a nonperturbative result.

\begin{theorem}[Ferromagnetism in a non-singular Hubbard model]
Suppose that the two dimensionless parameters $t/s$ and $U/s$ are 
sufficiently large.
Then the ground states have $\Stot=\Smax(=N/4)$ and are nondegenerate 
apart from the trivial spin degeneracy.
\label{Hal}
\end{theorem}

The theorem is valid, for example, 
when $t/s\ge4.5$ if $U/s=50$, and $t/s\ge2.6$ if 
$U/s=100$.
The ferromagnetic ground state can be constructed in exactly the same 
manner as $\Phi_{\up}$ in the previous section.

Although the model is rather artificial, this is the first rigorous 
example of saturated ferromagnetism in a non-singular \Hub\ on which we 
have to overcome the competition between $\Hint$ and $\Hhop$.
In a class of similar models, it is also proved that low-lying excitation 
above the ground state has a normal dispersion relation of a 
spin-wave excitation \cite{95c,95b}.
Starting from a \Hub\ model of itinerant electrons, the existence of a 
``healthy'' ferromagnetism is established rigorously.

If we set $s=0$ in the present model, the ground states of the 
single-electron \Sch\ equation become $N/2$-fold degenerate.
In this case, the model exhibits saturated ferromagnetism (flat-band 
ferromagnetism) for any $U>0$.
Theorem \ref{Hal} for $s\ne 0$ can be regarded as a solution to the 
problem about stability of flat-band ferromagnetism against perturbation 
to the hopping Hamiltonian.

The basic strategy in the proof of Theorem \ref{Hal} is first to 
establish the existence of saturated ferromagnetism in a \Hub\ on a chain 
with five sites, and then ``connect'' together these local ferromagnetism 
to get ferromagnetism in the whole system.
Generally speaking, this is a crazy idea.
In a quantum mechanical system, especially in a system with healthy 
dispersion (like the present one), electrons have strong tendency to 
extend in a large region and reduce kinetic energy.
To confine electrons in a finite region usually costs extra energy.
To obtain exact information about large system from smaller 
system seems to be impossible.
The reasons that such a strategy works in the present model are twofold.
One is the special construction of the model.
The other is that we described electron states using a language which 
takes into account both the particle-like character of electrons and the 
band structure of the model.
The latter is a natural strategy to deal with the \Hub\ in which 
wave-particle dualism generate interesting physics.

It is believed that the ground states of the present \Hub\ describes an 
insulator.
When the electron number is less than $N/2$, we expect that the present 
model exhibits metallic ferromagnetism in which the same set of electrons 
participate in conduction as well as magnetism\footnote{
By using a heuristic perturbation theory based on the Wannier functions, 
the low energy effective theory of the present \Hub\ is shown to be the 
ferromagnetic $t$-$J$ model.
Moreover, by considering models close to the flat-band model, one can make 
$|t/J|$ arbitrarily small.
This observation gives a strong support to the above conjecture.
}.
For the moment, we still do not know of any useful ideas in proving this 
fascinating conjecture.

\bigskip\bigskip
I wish to thank Kenn Kubo 
and Balint T\'{o}th for useful comments on 
the early version of the present article.
%%%%%%%%%%%%%%%%%%%%%%%%%%%%%%%%%%%%%%
%%%%%%%%%%%%%%%%%%%%%%%%%%%%%%%%%%%%%%
%%%%%%%%%%%%%%%%%%%%%%%%%%%%%%%%%%%%%%
%\bibliography{myWorks,CM,JAP}
%\bibliographystyle{prsty}

%\newpage

\end{document}